\documentclass[aps,pre,preprint,showpacs]{revtex4}
\usepackage{epsf}
\usepackage{amsmath,amssymb,enumerate,epsfig,color}

\begin{document}

\title {Decay Rate Statistics of Unstable Classically Chaotic Systems}

\author{Valentin V. Sokolov}
\affiliation{Budker Institute of Nuclear Physics, Novosibirsk, Russia}
\date{\today}

\begin{abstract}
Decay law of a complicated unstable state formed in a high energy collision
is described by the Fourier transform $K(t)$ of the two-point correlation 
function of the scattering matrix. Although each constituent resonance state
decays exponentially the decay of a state composed of a large number $N\gg 1$
of such interfering resonances is not, generally, exponential. We introduce the 
decay rates distribution function $w(\Gamma)$ by representing the decay law in 
the form of the mean-weighted decay exponent $K(t)=\int_0^{\infty}d\Gamma\,
e^{-\Gamma t}\,w(\Gamma)$. In the framework of the random matrix approach we 
investigate the properties of the distribution function $w(\Gamma)$ and its relation 
to the more conventional statistics of the decay widths. The latter is not in fact 
conclusive as concerns the evolution at the times shorter than the characteristic 
Heisenberg time. Exact analytical consideration is presented for the case of systems 
without time reversal symmetry.   

\end{abstract}
\maketitle

\section{Introduction}
The temporal aspect of chaotic resonance scattering is an interesting and important
issue which repeatedly attracted much attention starting from the seminal works of 
Wigner and Smith \cite{Wigner,Smith}. Later on different problems concerning the duration 
of resonance collisions were posed and discussed in detail. Distribution of resonance 
widths on the one hand and statistics of partial and proper delay times on the other 
have been investigated in a number of papers \cite{Porter,Lyuboshitz,Lewenkopf,Eckhard,
Dittes,Izrailev,Lehmann,Gopar,Fyodorov1,Fyodorov2,Brouwer,Kottos1,Sommers,Kottos2}.   

It is generally known that a single quasi-stationary state formed by the time 
$t=0$ in a collision with a given energy $E$ decays afterwards exponentially with 
the decay rate $\Gamma_r$ which is given by the imaginary part of the resonance 
complex energy ${\cal E}_r=E_r-\frac{i}{2}\Gamma_r$. In accordance with the Bohr 
energy-time uncertainty principle this rate defines the width of the Breit-Wigner
resonant curve.

However at high enough collision energy $E$ a very complicated unstable state is 
formed which is a superposition of many, $N\gg 1$, interfering resonance contributions 
whose spectrum is given by the $N$ complex eigenvalues 
${\cal E}_r =E_r-\frac{i}{2}\Gamma_r$ of $N\times N$ matrix of the non-Hermitian 
effective Hamiltonian $\,\,\,{\cal\hat H}={\hat H}-\frac{i}{2}{\hat W},\,\,\,
{\hat W}={\hat A}{\hat A}^{\dag}\,.\,\,\,$ The resonances, generally, overlap so 
that their mean width is larger than the mean level spacing, 
$\langle\Gamma\rangle > \Delta$. The evolution of such an unstable configuration 
formed via an incoming channel $a$ and decaying onto an outgoing channel $b$  
is described \cite{Lyuboshitz,Dittes} by the Fourier transform
\begin{equation}\label{DecLaw}
K^{b a}(\tau)=\frac{1}{2\pi}\int_{-\infty}^{\infty} d\omega\,
e^{i\omega\tau}\,C^{b a}(\omega)\,
\end{equation} 
of the two-point S-matrix correlation function
\begin{equation}\label{Scorr}
C^{b a}(\omega=t_H\varepsilon)=\frac{\langle S^{b a}(E+\varepsilon/2)\,
{S^{b a}}^*(E-\varepsilon/2)\rangle_{conn}}{T^b\,T^a}\,;\qquad b\neq
a\,
\end{equation}
Here $T^c=1-|S^{cc}|^2$ is the transmission coefficient in the channel $c$ and 
$t_H=2\pi/\Delta$ is the Heisenberg (Weisskopf) time. It is convenient to measure
the time in the units of this characteristic time interval, $\tau=t/t_H$. 
Correspondingly, the dimensionless energy shift $\omega=\varepsilon t_H$ is 
introduced in the Eqs. (\ref{DecLaw}, \ref{Scorr}). To avoid complications 
irrelevant in the present context we restrict for a while our consideration 
to inelastic collisions only. 

Contrary to the case of an isolated quasi-stationary state, the function 
$K^{b a}(\tau)$ does not, generally, decays exponentially and a connection to the 
widths of the constituting resonances is not immediately seen. More than that, 
an opinion has not once been expressed that the very notion of the resonance widths
becomes irrelevant when resonances strongly overlap. Nevertheless, intuitively, one 
would expect that decay properties of a complicated unstable state should somehow 
depend on the distribution of the individual resonance poles of the scattering 
amplitudes in the complex energy plane. Could then a non-exponential decay results 
from an incoherent mixture of exponential decays with different rates? If and how such 
a mixture is related to the statistics of the widths of resonances?  We address these 
questions below. They are answered explicitly by the example of the chaotic systems 
without time-reversal symmetry. A rigorous analytical solution is found in this case.       

\section{decay rates distribution function}
The function $C^{b a}(\omega)$ is analytical in the 
lower part of the complex $\omega$ plane and, obviously, satisfies the condition
$C^{b a}(-\omega)={C^{b a}}^*(\omega)\,.$ As a result its Fourier transform $K^{b
a}(\tau)\equiv 0$ when $\tau < 0$ and is real when $\tau > 0$. More than that, 
independently of the number $M$ of the reaction channels, \cite{Gorin}
\begin{equation}\label{K(0)=1}
K^{b a}(0)=1\,\,\,\, \mathrm{and}\,\,\,\, K^{b a}(\tau) > 0\,\,\,\,\,
\mathrm{when}\,\,\,\tau > 0\,.
\end{equation}
Notice at last that the mean reaction cross section is expressed as
\begin{equation}\label{sigma_ba}
\langle\sigma^{b a}\rangle=T^b\,T^a\,C^{b a}(0)=
T^b\,T^a\int_0^{\infty} d\tau\,K^{b a}(\tau)\,.
\end{equation}

Such properties suggest the Laplace representation for the decay law function
\begin{equation}\label{Laprep}
K^{b a}(\tau)=\int_0^{\infty} d\gamma\,w^{b a}(\gamma)\,
e^{-\gamma\tau}\,.
\end{equation}
(In the ordinary units
$\gamma=\Gamma\,t_H=\frac{2\pi}{\Delta}\,\Gamma$ where $\Gamma$ is the decay rate.) 
In view of the Eq. (\ref{K(0)=1}) the function $w^{b a}$ is normalized to unity,
\begin{equation}\label{Norm_w}
\int_0^{\infty} d\gamma\,w^{b a}(\gamma)=
K^{b a}(0)=1\,.
\end{equation}
It follows from the Eqs. (\ref{DecLaw}, \ref{Laprep}) that
\begin{equation}\label{Reciprep}
C^{b a}(\omega)=\int_0^{\infty} d\gamma\,
\frac{w^{b a}(\gamma)}{\gamma+i\omega}
\end{equation}
so that the correlation function is not, generally, single-valued in the complex 
$\omega$ plane. Finally, according           to the Eqs. (\ref{sigma_ba}, \ref{Reciprep}),
\begin{equation}\label{sigma_w}
\langle\sigma^{b a}\rangle= T^b\,T^a\int_0^{\infty}
d\gamma\frac{w^{b a}(\gamma)}{\gamma}\,.
\end{equation}
From this point on we suppose that all channels are
statistically equivalent, $T^b=T^a=T$, so that the functions
$C(\omega),\,K(\tau),\,w(\gamma)$ do not depend on the channel
indices.

\section{The semiclassical asymptotic expansion}

Let us consider first the semiclassical limit: $N\gg 1,\,\,M\gg 1$ but the ratio
$m=M/N\,,$ is finite though small, $m\ll 1,$ (this condition is physically justified). 
The function $w(\gamma)$ can easily be extracted from the results reported in: 
\cite{Lewenkopf,Lehmann} (GOE)
\begin{equation}\label{Verexp}
C(\omega)=\frac{1}{MT}\,\frac{\gamma(\omega)}{\gamma(\omega)+i\omega}\Rightarrow
\frac{1}{\gamma_W+i\omega}-\frac{2T}{(\gamma_W+i\omega)^2}+
\frac{MT^2}{(\gamma_W+i\omega)^3}+...
\end{equation}
where $\gamma(\omega)$ is a slow varying function and $\gamma(0)\equiv
\gamma_W=MT=t_H\Gamma_W$ where $\Gamma_W=\frac{\Delta}{2\pi}\,MT$ is the 
Weisskopf width. The asymptotic series in the r.h.s which is obtained by 
expanding near the pole $\omega=i\gamma_W$ coincides with that derived first
in different manner in: \cite{Verbaarschot}. The first term corresponds to 
the Hauser-Feshbach approximation and implies the purely exponential decay 
$e^{-\gamma_W\tau}$ when the subsequent ones account for deviations. In the 
limit $N, M\rightarrow\infty,\,\,0<m\ll 1$ the Weisskopf width equals to the 
empty gap between the real energy axis and the cloud of the poles corresponding 
to contributing resonances \cite{Lehmann}. The width $\Gamma_W$ is in this case 
the smallest possible width.  

It immediately follows from the expansion (\ref{Verexp}) that
\begin{equation}\label{Asymp_w}
w(\gamma)=\delta(\gamma-\gamma_W)-2T\,\delta'(\gamma-\gamma_W)+
\frac{MT^2}{2}\,\delta''(\gamma-\gamma_W)+...\,.
\end{equation}
The function $w(\gamma)$ thus obtained is neither smooth nor positive definite. 
We will see below that the expansion (\ref{Verexp}) is in this respect somewhat
misleading. Nevertheless it works well in certain cases. In particular, for 
the mean cross section we get from the Eqs. (\ref{sigma_w}, \ref{Asymp_w})
\begin{equation}\label{sigmaexp}
\langle\sigma\rangle^{(GOE)}=\frac{T}{M}\left(1-\frac{1}{M}+...\right)\,.
\end{equation}
Similar calculation yields in the GUE-case
$\,\,\,\langle\sigma\rangle^{(GUE)}=\frac{T}{M}\,.$

Notice that the difference
$\langle\sigma\rangle^{(GUE)}-\langle\sigma\rangle^{(GOE)}=\frac{T}{M^2}$
determines the "weak localization" in the quantum transport.

\section{decay rates versus widths statistics}
Heuristic" arguments have been adduced in \cite{Dittes91} which assume 
a simple connection between the function $w(\gamma)$ and the distribution 
$\rho_M(\gamma)$ of the resonance widths
\begin{equation}\label{Intu_w_rho}
w_M(\gamma)=A\,\gamma^2\,\rho_M(\gamma)
\end{equation}
with some normalization constant $A$. A formula of such a kind has been derived a 
little later in \cite{Izrailev}. It was shown, however, that
the relation supposed can reproduce only the long-time asymptotic behavior. Only 
the most long-lived resonances survive by this time, so they can be expected to obey 
the $\chi^2_M$ widths statistics. 
\begin{equation}\label{chi_stat}
\rho_M(\gamma)=\frac{1}{\,\left(\beta\frac{M}{2}\right)!}\,\frac{1}{\gamma}\,
\left(\beta\frac{\gamma}{2T}\right)^{\beta\frac{M}{2}}\,
e^{-\beta\,\frac{\gamma}{2T}}, \quad \langle\gamma\rangle=MT=\gamma_W,\,\,\,
\beta=1,\, 2 \,.
\end{equation}
Here the parameter $\beta$ marks the Dyson's symmetry class: GOE ($\beta=1$) or GUE 
($\beta=2$). Notice that in our dimensionless units $\Gamma/\langle\Gamma\rangle=
\gamma/MT$ which means that $\langle\gamma\rangle=MT=\gamma_W$. In the case of 
isolated resonances the Weisskopf width defines the mean rather than the minimal width. 
The formal extending the relation (\ref{Intu_w_rho}) to all values of $\gamma$ yields 
the result
\begin{equation}\label{Intu_K}
K_M(\tau)=A\,M\left(M+\frac{\beta}{2}\right)
\frac{T^2}{\left(1+\frac{2T}{\beta}\tau\right)^{\beta\frac{M}{2}+2}}
\rightarrow\frac{const.}{\tau^{\beta\frac{M}{2}+2}}
\end{equation}
which reproduces rightly the power law asymptotic behavior \cite{Lewenkopf} though 
the proper calculation of the constant is beyond the validity of this approximation.
The constant $A$ can be chosen to quantitatively fit the correct asymptotics. However 
at shorter times the found in such a way expression disagree with the actual decay law, 
the discrepancy being the stronger the closer is the transmission coefficient $T$ to 
its maximal value 1 (see \cite{Izrailev}). There is no way to reconcile all necessary 
conditions which must me satisfied by the decay function $K_M(\tau)$ by means of only one
matching parameter $A$. There are two possible reasons why the supposed relation (\ref{Intu_w_rho}) fails: i. the $\chi_M^2$ distribution is not valid when resonances 
overlap; ii. the naive derivation is wrong. We will clarify below these guess-work by the example of the systems with broken time-reversal symmetry which are described by the
Gaussian unitary ensemble (GUE) of random Hamiltonians.

\section{Systems with no time-reversal symmetry}
\subsection{General consideration}
 The GUE two-point correlation function derived in: \cite{Savin1} reduces in the case of equivalent channels to 
\begin{equation}\label{C_GUE}
\begin{array}{c}
C(\omega)=\int_0^1 d\lambda\int_0^{\infty} d\lambda_1\,
e^{-i\omega(\lambda+\lambda_1)}\,
\frac{1}{\lambda+\lambda_1}
\frac{(1-T\lambda)^{M-2}}{(1+T\lambda_1)^{M+2}}\,\times\\
\left\{\left(\frac{2}{T}-1\right)(1-T\lambda)(1+T\lambda_1)-
\frac{1-T}{T}\left[(1-T\lambda)+(1+T\lambda_1)\right]\right\}\,.\\
\end{array}
\end{equation}
Making use of the following simple identity
\begin{equation}\label{Id1}
\frac{1}{\lambda_1+\lambda}=\int_0^{\infty}
d\gamma\,e^{-\gamma(\lambda_1+\lambda)}
\end{equation}
we factorize the integrations over $\lambda$ and $\lambda_1$. On the next step we use
a second identity
\begin{equation}\label{Id2}
\int_0^{\infty} d\lambda_1\frac{e^{-(\gamma+i\omega)\lambda_1}}{(1+T\lambda_1)^{M+1+\mu}}=
\frac{1}{(M+\mu)!}\int_0^{\infty} d\eta \,
\frac{\eta^{M+\mu}\,e^{-\eta}}{T\eta+\gamma+i\omega},\,\,\mu=0, 1\,.
\end{equation}
The correlation function (\ref{C_GUE}) is thus reduced to a linear combination of  
terms like 
\begin{equation}\label{Par_c}
c_{\mu\mu'}(\omega)=\frac{1}{(M+\mu)!}\int_0^{\infty} d\gamma\int_0^{\infty} d\eta\,
\eta^{M+\mu}\,e^{-\eta}\int_0^1 d\lambda (1-T\lambda)^{M-1-\mu'}\,e^{-\gamma\lambda}
\frac{e^{-i\omega\lambda}}{T\eta+\gamma+i\omega}\,.
\end{equation} 
After a chain of simple transformations of the integration variables the Fourier transform
of the latter function reduces finally to
\begin{equation}\label{Par_k}
\begin{array}{c}
k_{\mu\mu'}(\tau)=\frac{1}{T^{2+\mu+\mu'}(M+\mu)!}\int_0^{\infty} 
d\gamma\,e^{-\gamma\tau}
\int_0^{\gamma} d\gamma'\,\gamma'^{(\mu+\mu')}\times\\
\int_{\frac{1-T}{T}\gamma'}^{\gamma'/T}
d\nu\, \Theta\left(\tau-\frac{1}{T}+\frac{\nu}{\gamma'}\right)\,
\nu^{M-1-\mu'}\,e^{-\nu}\,.\\
\end{array}
\end{equation}  
The symbol $\Theta$ stands for the step function. All successive integrations in 
this expression can be carried out explicitly \cite{Richter}. 

To simplify further consideration we restrict our calculation to the 
case of the perfect coupling to the continuum, $T=1$, when the naive 
consideration is the least satisfactory. Only the term with $\mu=\mu'=0$ 
remains in this limit and
\begin{equation}\label{Perf_C}
C_M(\omega)=\frac{1}{M!}\int_0^{\infty}d\gamma\int_0^{\infty}d\gamma'\,\gamma'^M\,
e^{-\gamma'}\int_0^1d\lambda\,(1-\lambda)^{M-1}\,e^{-\gamma\lambda}\,
\frac{e^{-i\omega\lambda}}{\gamma+\gamma'+i\omega}\,.
\end{equation}
The subscript $M$ explicitly indicates the number of open channels. Notice that when
$T=1$ the formula (\ref{Perf_C}) and so all consequent results are valid also for the 
elastic collisions including the purely elastic process with one open channel $M=1$. 

The found representation enables us to evaluate the mean cross section in the following 
simple and elegant way: 
\begin{equation}\label{meansigma}
\langle\sigma\rangle=C_M(0)=\frac{1}{M!}\int_0^{\infty}d\gamma\,
F_M(\gamma)\,
\Phi_{M-1}(\gamma)=\frac{1}{M}\,\int_0^{\infty}d\gamma\,
e^{-\gamma}\,F_0(\gamma)
=\frac{1}{M}\,.
\end{equation} 
The third equality follows from the fact that the functions
\begin{equation}\label{Funcs}
F_M(\gamma)=\int_0^{\infty}d\eta\,\eta^M\,\frac{e^{-\gamma\eta}}{1+\eta},
\qquad\Phi_{M-1}(\gamma)=\int_0^{\gamma}d\zeta\,
(\gamma-\zeta)^{M-1}\,e^{-\zeta}
\end{equation}
which appear in this equation obey the simple recursions
\begin{equation}\label{Recurs}
F_M(\gamma)=-\frac{d}{d\gamma}\,F_{M-1}(\gamma),
\qquad \frac{d}{d\gamma}\,\Phi_M(\gamma)=M\,\Phi_{M-1}(\gamma)\,.
\end{equation}

The decay function looks now as
\begin{equation}\label{Perf_K}
K_M(\tau)=\int_0^{\infty}d\gamma\,e^{-\gamma\tau}\,\frac{1}{M!}\,
\int_0^{\gamma}d\gamma'
\,\int_0^{\gamma'}d\nu\,\Theta(\tau-1+\nu/\gamma')\,\nu^{M-1}\,e^{-\nu}\,.
\end{equation}
The later consideration depends on whether the time $\tau$ exceeds the Heisenberg time 
$\tau_H=1$ or not.

\subsection{Long-time asymptotics}
After the Heisnberg time, $\underline{\tau > 1}\,,$ the step function in the Eq. (\ref{Perf_K}) equals to one in the whole integration region and the decay function gets 
the required form
\begin{equation}\label{K_after}
K_M(\tau>1)=
\int_0^{\infty}d\gamma\,e^{-\gamma\tau}\,w_{>}(M,\gamma)
\end{equation}
with the decay rates distribution function given by
\begin{equation}\label{w_after}
\begin{array}{c}
w_{>}(M,\gamma)=\frac{1}{M!}\int_0^{\gamma} d\gamma'\varphi_{M-1}(\gamma'),\\
\varphi_{M-1}(\gamma)=\int_0^{\gamma} d\nu\,\nu^{M-1}\,e^{-\nu}=
(M-1)!-\Gamma(M,\gamma)\\
\end{array}
\end{equation}
where $\Gamma(M,\gamma)$ is the incomplete $\Gamma$ function
\begin{equation}\label{Inc_Gamma}
\Gamma(M,\gamma)=(M-1)!\,e^{-\gamma}\sum_{m=0}^{M-1}\,
\frac{\gamma^m}{m!}\,.
\end{equation}
The next integration over $\gamma'$ yields finally
\begin{equation}\label{w_after1}
\begin{array}{c}
w_{>}(M,\gamma)=\frac{\Gamma(M+1,\gamma)-\gamma\,\Gamma(M,\gamma)}{M!}+
\gamma\frac{1}{M}-1=\widetilde{w}_M(\gamma)-\left(1-\gamma\frac{1}{M}\right),\\
\widetilde{w}_M(\gamma)=e^{-\gamma}\sum_{m=0}^{M-1}\,
\left(1-\frac{m}{M}\right)\,
\frac{\gamma^m}{m!}\,\equiv e^{-\gamma}\,P_{(M-1)}(\gamma)\,.\\
\end{array}
\end{equation}
As expected, this distribution is smooth and positive definite. Notice that, due to 
subtraction of the last two terms, the distribution $w_{>}(M,\gamma)$ vanishes as 
$\gamma^2$ when $\gamma\rightarrow 0$.

At last the final integration over $\gamma$ yields  
\begin{equation}\label{K_after_expl}
K_M(\tau>1)=\int_0^{\infty} d\gamma\,e^{-\gamma\tau}\,w_{>}(M,\gamma)=
\frac{1}{M}\,\frac{1}{\tau^2\,(1+\tau)^M}\,, \quad
K_M(\tau\rightarrow\infty)
=\frac{\langle\sigma\rangle}{\tau^{(M+2)}}
\end{equation}\\
thus fixing not only the power but also the constant of asymptotic power behavior. 

The found results are closely connected to statistics of the decay width. Indeed the 
widths distribution function in the case of $M$ equivalent channels, arbitrary degree of 
resonance overlapping and perfect coupling to the continuum is currently well known to be 
\cite{Fyodorov1}
\begin{equation}\label{Dwidth}
\rho_{M}(\gamma)=\frac{1}{(M-1)!\,
\gamma^2}\int_0^{\gamma} d\nu\,\nu^M\,e^{-\nu}\,
\end{equation}
and has nothing to do with the $\chi^2_M$-distribution (\ref{chi_stat})
It is immediately seen that
\begin{equation}\label{phy_rho}
\varphi_{M-1}(\gamma)=(M-2)!\,\gamma^2\,\rho_{M-1}(\gamma)\,
\end{equation}
(see the Eq. (\ref{w_after})). 
This results in the following connection between the two distributions
\begin{equation}\label{Drates_Width}
\begin{array}{c}
w_{>}(M,\gamma)=\frac{1}{M(M-1)}\int_0^{\gamma}
d\gamma'\,\gamma'^2\,
\rho_{M-1}(\gamma')=\\
\frac{1}{M^2}\left[\gamma^2\,\rho_{M}(\gamma)+\int_0^{\gamma}
d\gamma'\,\gamma'^2\,\rho_{M}(\gamma')\right]=
\frac{1}{M^2}\left(\frac{d}{d\gamma}+1\right)
\int_0^{\gamma} d\gamma'\,\gamma'^2\,\rho_{M}(\gamma')\,.\\
\end{array}
\end{equation}
The extra term which appears in this relation is missing in the naive 
formula (\ref{Intu_w_rho}).

\subsection{Short-time behavior}
At the times shorter than $\tau_H$,$\,\,\,\underline{\tau < 1}\,,$
an additional term arises,
\begin{equation}\label{}
\int_0^{\gamma}d\nu\,\Theta(\tau-1+\nu/\gamma)\,\nu^{M-1}\,e^{-\nu}=
\varphi_{M-1}(\gamma)
-\int_0^{(1-\tau)\gamma}d\nu\,\nu^{M-1}\,e^{-\nu}\,,
\end{equation}
which depends on the time. As a consequence
the decay law contains along with the smoothly distributed exponential 
contributions an additional polynomial term
\begin{equation}\label{K_before}
\begin{array}{c}
K_M(\tau<1)=\int_0^{\infty} d\gamma\,e^{-\gamma\tau}\,
\widetilde{w}_M(\gamma)+
\frac{1}{M}\,\frac{1-(1-\tau)^M}{\tau^2}-\frac{1}{\tau}=\\
\int_0^{\infty}
d\gamma\,e^{-\gamma\tau}\,\widetilde{w}_M(\gamma)-
\frac{1}{M}\sum_{m=0}^{M-2}(-1)^m\, \left(
\begin{array}{c}
M\\
m+2\\
\end{array}
\right)\,\tau^m\,.\\
\end{array}
\end{equation}
Formally, we still can present this formula in the form
\begin{equation}\label{FK_before}
K_M(\tau<1)=\int_0^{\infty} d\gamma\,e^{-\gamma\tau}\,w_{<}(M,\gamma)
\end{equation}
with a "distribution density" 
$w_{<}(M,\gamma)=\widetilde{w}_M(\gamma)+w_M^{(sing.)}(\gamma)\,.$ The price 
paid is the singular nature of the additional weight function 
\begin{equation}\label{w_sing}
w_M^{(sing.)}(\gamma)=-\frac{1}{M}\sum_{m=0}^{M-2}(-1)^m\,
\left(
\begin{array}{c}
M\\
m+2\\
\end{array}
\right)\,\frac{d^m}{d\gamma^m}\,\delta(\gamma),\quad M\geqslant 2\,.
\end{equation} 
This singular "distribution" differs in two important respects from that extracted 
from the Verbaarschot's asymptotic expansion: first, it contains only a 
finite sum of derivatives of the $\delta$-function and, second, all singular 
terms are concentrated near $\gamma=0$ rather than $\gamma=\gamma_W$.   

We should stress that the both regular and singular parts play, generally, equally
important roles. In particular,
\begin{equation}\label{K_0_b+a}
\quad K_M(0)=\frac{M+1}{2}-\frac{M-1}{2}=1\,.
\end{equation}
The first contribution comes from the regular and the second from the singular 
parts of the "distribution" $w_{<}(M,\gamma)$. They are of the same order of magnitude 
when $M\gg 1$.

Integration over $\gamma$ gives now
\begin{equation}\label{K_berore_expl}
K_M(\tau<1)=\frac{1}{M}\,\frac{1-(1-\tau^2)^M}{\tau^2(1+\tau)^M}\,,
\quad K_M(0)=1\,.\\
\end{equation}
Combining all found results we arrive finally at
\begin{equation}\label{K_expl}
K_M(\tau)=\frac{1}{M}\,\frac{1}{\tau^2(1+\tau)^M}\,
\left[1-\Theta(1-\tau)(1-\tau^2)^M\right]\,,\quad
0\leqslant\tau<\infty\,.
\end{equation} 
This function satisfies all necessary conditions and is continuous though not
analytical in the point $\tau=\tau_W=1$.

In the semiclassical limit $M\gg 1$ the characteristic decay time
$\tau_W=1/\gamma_W=1/M\ll 1$ is much shorter than the Heisenberg time.
The polynomial term is therefore of principal importance. It becames 
obvious from the following equivalent presentation of the 
Eq. (\ref{K_berore_expl})
\begin{equation}\label{AsK_before}
\begin{array}{c}
K_M(\tau<1)=\frac{1}{M\tau^2}
\left[e^{-M\ln(1+\tau)}-e^{M\ln(1-\tau)}\right]=\\
e^{-M(\tau+O(\tau^3))}\,
\frac{2\sinh\left[\frac{M}{2}(\tau^2+O(\tau^4))\right]}{M\tau^2}
\approx e^{-M\tau}=e^{-\gamma_W\tau}\,.\\
\end{array}
\end{equation} 
Thus such an exponential semiclassical decay with the characteristic Weisskopf's 
decay rate $\gamma_W$ cannot be directly traced to the statistics of resonance 
widths. Deviation from the exponential law due to the neglected terms becomes 
significant after the time $\tau_q\sim1/\sqrt{M}\gg\tau_W$ \cite{Casati,Savin2}.

\subsection{Mean cross section}
The mean cross section can be expressed in two equivalent forms
\begin{equation}\label{sigms_K_w}
\begin{array}{c}
\langle\sigma\rangle=\frac{1}{M}=\int_0^1 d\tau\,K_M(\tau)+
\int_1^{\infty} d\tau\,K_M(\tau)=\\
\int_0^{\infty}\,\frac{d\gamma}{\gamma}\,
e^{-\gamma}\left[P_{(M-1)}(\gamma)-
1+\frac{\gamma}{M}\right]+\int_0^{\infty}\,w_M^{(sing.)}(\gamma)\,
\frac{1-e^{-\gamma}}{\gamma}\,.\\
\end{array}
\end{equation}
The long-time contribution ($\tau>1$) rapidly decreases when the number 
of channels $M$ grows. Even for $M=1, 2$ it amounts only to 30\% and 10\% 
respectively. When $M\gg 1$ the following asymptotic expansion
\begin{equation}\label{K_asym}
\int_1^{\infty} d\tau\,K_M(\tau)=K_M(1)\,\frac{2}{M}\int_0^{M/2} d\xi\,
e^{-(\xi+3\xi^2/2M+...)}\approx \frac{2}{M^2}\,e^{-M\ln2}
\end{equation}
shows that this contribution diminishes very fast. The mean cross section
is defined by the short time evolution. 

On the other hand, the contributions of the regular and singular parts of 
the decay rates distribution are equally important. Indeed
\begin{equation}\label{Sigma_smooth}
\int_0^{\infty}\,\frac{d\gamma}{\gamma}\,
e^{-\gamma}\left[P_{(M-1)}(\gamma)-
1+\frac{\gamma}{M}\right]=\frac{1}{M}+\sum_{m=2}^M\frac{1}{m}\,.
\end{equation}
The logarithmically growing extra term is perfectly compensated 
by the contribution of the singular part.

\section{Conclusions}
The decay law $K(t)$ of a complicated unstable state of a classically chaotic 
system can be formally presented in the form of a Laplace integral as a weighted 
mean value of the decay exponents. By this the notion of the decay rates distribution 
is introduced irrelative of statistics of the resonance poles of the scattering 
amplitudes. The connection of this distribution with statistics of the resonance 
widths is then investigated. Exact analytically solution is found in the case of systems 
with broken time-reversal symmetry. It is demonstrated that only the long-time 
$\tau\gg\tau_H=1$ asymptotic behavior of the decay law is governed by the statistics
of the resonance widths. For the shorter times $\tau<\tau_H=1$ the contributions of the 
connected to the widths statistics smooth part and the singular part of the "distribution" $w_{<}(M,\gamma)$ are equally important. In particular, the approximately exponential semiclassical decay with the characteristic Weisskopf's decay rate $\gamma_W$ results 
from interrelation of the both of them.

\section{acknowledgements}
I am very grateful to D.V. Savin for his interest to this work, plentiful discussions and advice. I also greately appreciate discussion of the results with V. Zelevinsky and A. Richter. The financial support from the RAS Joint scientific program "Nonlinear dynamics and solitons" is acknowledged with thanks.


\begin{thebibliography}{99}
\bibitem{Richter} Explicit calculation of the (GUE) decay function $K(t)$ has been
fulfilled by prof. A. Richter and his coworkers. Privete communication.
\end{thebibliography}
\end{document}